\documentclass[pra,aps,twocolumn,floatfix,superscriptaddress,showpacs]{revtex4}
\usepackage{graphicx}
\usepackage{amsmath}

\newcommand{\eq}{\begin{equation}}
\newcommand{\fine}{\end{equation}}

\begin{document}

\title{

\bf \LARGE  Pomeron pole plus grey disk model :\\
real parts, inelastic cross sections and LHC data
}

\author{S. M. Roy}
\email{smroy@hbcse.tifr.res.in} \affiliation{HBCSE,Tata Institute of Fundamental Research, Mumbai}

\begin{abstract}
I propose a two component analytic formula $F(s,t)=F^{(1)}(s,t)+F^{(2)}(s,t)$ for 
$(ab\rightarrow ab) +(a\bar{b}\rightarrow a\bar{b})$ scattering at energies $\ge 100 GeV$ ,where $s,t$ denote 
squares of c.m. energy and momentum transfer.It saturates the Froissart-Martin bound and obeys 
Auberson-Kinoshita-Martin (AKM) \cite{AKM1971} \cite{Auberson-Roy} scaling. I choose 
$Im F^{(1)}(s,0)+Im F^{(2)}(s,0)$ as given by Particle Data Group (PDG) fits \cite{PDG2005},\cite{PDG2013} to 
total cross sections, corresponding to simple and triple poles in angular momentum plane. The PDG formula is extended to 
non-zero momentum transfers using partial waves of $Im F^{(1)}$ and $Im F^{(2)}$ motivated by Pomeron pole and 
'grey disk' amplitudes and constrained by inelastic unitarity. $Re F(s,t)$ is deduced  from real analyticity: 
I prove that 
$Re F(s,t)/ImF(s,0)  \rightarrow (\pi/\ln{s})  d/d\tau (\tau  Im F(s,t)/ImF(s,0) )$ for $s\rightarrow \infty$ with 
$\tau=t (ln s)^2$ fixed, and apply it to $F^{(2)}$.Using also the forward slope fit by Schegelsky-Ryskin \cite{Schegelsky-Ryskin},
the model gives real parts,differential cross sections for $(-t)<.3 GeV^2$, and inelastic cross sections in good agreement with data 
at $546 GeV, 1.8 TeV,7 TeV$ and $ 8 TeV $. 
It predicts for inelastic cross sections for $pp$ or $\bar{p} p$, 
$\sigma_{inel}=72.7\pm 1.0\> mb$ at $7TeV$ and $74.2 \pm 1.0\>mb$ at $8 TeV$ in agreement with $pp$ 
Totem \cite{Totem1}\cite{Totem2}
   \cite{Totem3}\cite{Totem4} experimental values $73.1\pm 1.3 mb $ and $74.7\pm 1.7 mb$ respectively, and with Atlas \cite{Atlas1}\cite{Atlas2}\cite{Atlas3}\cite{Atlas4}
  values $71.3\pm 0.9\> mb$  and $71.7\pm 0.7\>mb$ respectively. The predictions $\sigma_{inel}=48.1\pm 0.7\> mb$ at $546 GeV$ and $58.5 \pm 0.8\>mb$ at $1800 GeV$ also 
  agree with $\bar{p} p$ experimental results of Abe et al \cite{Abe}  $48.4\pm .98 mb $ at $546 GeV$ and $60.3\pm 2.4 mb$ at $1800 GeV$. 
  The model yields for 
$\sqrt{s}> 0.5 TeV$, with  PDG2013 \cite{PDG2013} total cross sections , and Schegelsky-Ryskin slopes \cite{Schegelsky-Ryskin} as input, $\sigma_{inel} (s) =
 22.6 + .034 ln s + .158 (ln s)^2 mb ,\> and 
\sigma_{inel} / \sigma_{tot} \rightarrow  0.56,\> s\rightarrow \infty ,$ where $s$ is in $GeV^2$ units.Continuation to positive $t$ indicates an 
'effective' $t$-channel singularity at  $\sim (1.5 GeV)^2$ ,and suggests that usual Froissart-Martin bounds are quantitatively weak as 
they only assume absence of singularities upto $4 m_\pi ^2$ .
\end{abstract}


\maketitle

% May skip begin and end figure options for lectures
% Caption works in the (begin and end) figure environment only
%\begin{figure}[ht]
%\begin{center}
% Use LaTeX and DVItoPDF while using EPS files
%\includegraphics[width=.75\columnwidth]{position_density.eps}
% Use PDFLaTeX while using JPG, PNG, PDF image files
%\includegraphics[width=.75\columnwidth]{W3.jpg}
%\caption{Position density}
%\label{fig:pos_dens}
%\end{center}
%\end{figure}

{\bf Introduction.} Precision measurements of  $pp$ cross sections at LHC 
 \cite{Totem1} \cite{Totem2}\cite{Totem3}\cite{Totem4}\cite{CMS}\cite{Atlas1}\cite{Atlas2}\cite{Atlas3}\cite{Atlas4}\cite{Alice}, 
 and in cosmic rays  \cite{Pierre Auger} 
motivate me to present a model for $ab\rightarrow ab$ 
scattering amplitude at c.m. energies $\sqrt{s} > 100 GeV$ described by an analytic formula containing very few parameters.
Neglecting terms with a power decrease at high $s$ , the Particle Data Group (PDG) fits  
to total cross sections \cite{PDG2005},\cite{PDG2013} are the sum of one constant component
and another rising as $(ln s )^2$ , corresponding to  a simple pole 
and a triple pole at $J=1$ in the angular momentum plane,
\begin{eqnarray}
&& \sigma^{ab }_{tot} =\sigma ^{(1),ab}_{tot}+\sigma ^{(2),ab}_{tot},\nonumber \\
&& \sigma ^{(1),ab}_{tot}=P^{ab},\> \sigma ^{(2),ab}_{tot}=H (\ln {s/s^{ab}_M } )^2 . 
\end{eqnarray}
I propose that, analogously, the full amplitude 
$F(s,t)=F^{(1)}(s,t)+F^{(2)}(s,t)$,  where, $F^{(1)}$ is a Pomeron simple pole amplitude , 
 $Im F^{(2)}$ has partial waves with a smooth cut-off at impact parameter $b=R(s)$   
corresponding to a grey disk and  $Re F^{(2)}(s,t)$ is calculated from a theorem I prove using real analyticity and  
Auberson-Kinoshita-Martin (AKM) \cite{AKM1971} \cite{Auberson-Roy} scaling for $s\rightarrow \infty$ with fixed $t (ln s)^2$. 
Inelastic  unitarity is tested using inputs of total cross sections, forward slopes and Pomeron parameters. Only inputs leading to unitary amplitudes 
are accepted. Model predictions for  inelastic cross sections,near forward real parts and differential cross sections agree with existing data and can be  
tested against future LHC experiments.

{\bf  Froissart-Martin bound basics.} Froissart \cite{Froissart1961},from the Mandelstam representation, and Martin \cite{Martin1966}, 
from axiomatic field theory, proved that the total cross-section $\sigma_{tot} (s)$ for 
two particles $a,b$ to go to anything must obey the bound,
\begin{equation}
\sigma_{tot} (s) \leq_{s\rightarrow \infty} C \> [\ln (s/s_0)]^2 ,
\end{equation}
where $C, s_0$ are unknown constants.It was proved later \cite{Lukaszuk} that $C=4\pi /(t_0)$, where $t_0$ is the lowest singularity 
in the $t-$channel .This 
bound has been extremely useful in theoretical investigations \cite{Roy1972} \cite{Kupsch} and  high energy models \cite{Cheng-Wu1970}
\cite{BSW1}\cite{BSW2}\cite{BSW3}\cite{BSW4}\cite{BSW5}\cite{Block1}\cite{Block2}\cite{Block3}\cite{Islam}. 
Analogous bounds on the inelastic cross section  have been obtained by Martin \cite{Martin2009}and Wu et al\cite{WMRS}; for 
pion-pion case, Martin and Roy  obtained bounds on energy averaged total\cite{Martin-Roy2014} and inelastic cross sections 
\cite{Martin-Roy2015} which also fix the scale factor $s_0 $ in these bounds.

{\bf Normalization.}For the $ab \rightarrow ab $ scattering amplitude $F(s,t)$, $a\neq b$, with $k=$ c.m. momentum, and $z=1 +t/(2k^2)$,
\begin{eqnarray}
 F(s,t)=\sqrt{s}/(4k) \sum_{l=0}^{\infty} (2l+1)P_l (z)  a_l (s), \nonumber \\
\sigma _{tot} (s) =4 \pi/(k^2) \sum _{l=0}^{\infty} (2l+1) Im a_l(s)\nonumber \\
\frac{d \sigma }{d t} = \frac{\pi} {k^2} \frac{d \sigma }{d \Omega } (s,t) = \frac {\pi} {k^2} \bigl |4 \frac{F(s,t)} {\sqrt{s} } \bigr |^2 .
\end{eqnarray}
with the inelastic unitarity constraint  $Im a_l (s) \geq | a_l (s) |^2 $. 
For identical particles $a=b$, the partial waves $ a_l (s) \rightarrow 2 a_l (s) $ in the above partial wave expansions for $F(s,t)$ , 
and $  \sigma _{tot} (s)$, but the odd partial waves are zero. We have the same formulae for the unitarity constraint, and 
the differential cross section as given above.

At high energy, using $a_l (s) \equiv  a(b,s)$, $ l= bk $, where $b$ is the impact parameter, and
$P_l ( cos \theta )\sim J_0 \big( (2l+1) \sin (\theta /2) \big)+O(  \sin ^2(\theta /2)), $ 
we have the impact parameter representaion,
\begin{eqnarray}
\label{Bessel}
&& F(s,t)=k \sqrt{s}/2 \int _0 ^{\infty} b db  a(b,s) J_0(b\sqrt{-t} )\nonumber\\ 
&&\sigma_{tot}= 8 \pi \int _0 ^{\infty} b db  Im a(b,s);\>\sigma_{el}= 8 \pi \int _0 ^{\infty} b db |a(b,s)|^2 \nonumber\\
&& d \sigma /d t =  4 \pi \big |\int _0 ^{\infty} b db  a(b,s) J_0(b\sqrt{-t} ) \big |^2, 
\end{eqnarray}

There exist very good fits to high energy data \cite{AGN} \cite{Martynov} with a very large number of free parameters . There are also very  
good eikonal based models involving several free parameters \cite{Cheng-Wu1970}
\cite{BSW1}\cite{BSW2}\cite{BSW3}\cite{BSW4}\cite{BSW5}\cite{Block1}\cite{Block2}\cite{Block3}\cite{Islam}. The recent eikonal based 
model of Block and Halzen (BH)\cite{BH1}\cite{BH2} uses high energy data to guess the glue-ball mass  and to probe whether the proton is a black disk.

{\bf A two component partial wave model.} I present a two component model with very few parameters and with analytic formulae 
for the total amplitude incorporating unitarity-analyticity constraints , PDG total cross sections and the 
AKM scaling theorem .

{\bf Imaginary parts.} I use the two component PDG total cross section fit. 
 I propose that in the impact parameter picture, the Imaginary part $Im a(b,s)$ of the partial waves at  fixed $s$ is also a sum of two components,  
 one part $Im a^{(1)} (b,s)$ a Gaussian corresponding to a Pomeron pole, and the other  $Im a^{(2)} (b,s)$ a polynomial of degree $2n$ in $b^2$ 
 with a smooth cut-off at $b=R(s)$ , $n$ being a positive integer. so that $Im a^{(2)} (b,s)$  is continuous and has continuous derivative at 
 $b=R(s)$. The second component corresponds to a ``grey'' disk with cross section rising as $(\ln {s})^2 $,
 \begin{eqnarray}
 \label{n}
& Im a (b,s) =Im a^{(1)} (b,s)  +Im a^{(2)} (b,s),\nonumber \\
&Im a^{(1)} (b,s)=C(s) \exp { (-2b^2 /D^2 (s))} ,\nonumber \\
&Im a^{(2)} (b,s)= E(s) (1-b^2/R^2(s))^{2n} \theta (R(s)-b), 
\end{eqnarray}
where $\theta(x)=1,\>for\>x \geq 0$, and $0$ otherwise.
The unitarity constraints are,
\begin{equation}
\label{unitarity constraints}
 C(s) \ge 0,\>E(s) \ge 0,\>0 \le C(s)+E(s) \le 1 \>.
\end{equation}
In Eq.(\ref{n}) we take the simplest choice $n=1$ in this paper.
Using the ansatz for $Im a^{(1)} (b,s)$, integrating over $b$ , and matching the result for $Im F^{(1)}(s,t)$ with the standard small $t$ 
Pomeron amplitude ,
\begin{equation}
\label{Pomeron}
 F^{(1)}(s,t)=\frac{k \sqrt{s} } {16\pi } \sigma ^{(1)}_{tot} \exp{(t b_P +t\alpha ' \ln {s }  ) }(i+t\frac{\pi}{2} \alpha '),
\end{equation}
we obtain ,
\begin{equation}
\label{CD^2}
 D^2(s)=8 ( b_P +\alpha ' \ln {s } ),\>C(s)=\sigma ^{(1)}_{tot} /(2\pi D^2 (s)).
\end{equation}
 Since $\sigma ^{(1)}_{tot} $ is a constant, $C(s)\rightarrow const/(\ln s),\> s\rightarrow \infty $ for $\alpha '\neq 0 $.
Similarly, the ansatz for $Im a^{(2)} (b,s)$ with $n=1$ yields,
\begin{equation}
 Im F^{(2)}(s,t) =E(s) \frac{4 k \sqrt{s} }{q^3 R(s) } J_3 (q R(s)),\>q\equiv \sqrt{-t},
\end{equation}
where $J_m(x)$ denotes the Bessel function of order $m$. Hence,
\begin{equation}
\label{ER^2}
\sigma ^{(2)}_{tot} (s)= \frac{16 \pi}{k \sqrt{s}} Im F^{(2)}(s,0)= \frac{4 \pi}{3} E(s)R^2 (s).
\end{equation}

Thus, $C(s)D^2(s)$ and $E(s)R^2(s)$ are determined from the PDF total cross section fits using Eqns.(\ref{CD^2}) and (\ref{ER^2}) respectively.
 A nice feature of the model is that the above unitarity constraints (\ref{unitarity constraints}) as well as a stronger version 
 including real parts can be readily tested, and provide acceptability criteria for extrapolations of experimental data for $pp$ scattering.
 
{\bf Theorem on Real parts.}Let $F(s,t)=F(y;t),\>y\equiv ((s-u)/2)^2$ be an $s-u$ symmetric amplitude, with asymptotic behaviour 
$|s| (\ln{|s|})^\gamma |\phi (\tau)|,\> \tau \equiv  t (\ln{|s/s_0|})^\beta $,
  where $\phi()$ is a real analytic function of it's argument and $\phi(0)=1$.
For fixed physical $t$, $F$ is real analytic in the cut-$y$ plane with only a right-hand cut from $(2m_a m_b +t/2)^2 $ to $\infty$. $F$ must be 
 real for $y= |y|\exp {(i\pi)},\>i.e. s \rightarrow |s|\exp {(i\pi /2)}$, and hence replacing $|s|\rightarrow s\exp {(-i\pi /2)}$, we have 
 for $s\rightarrow \infty\>,\tau \> fixed $,
 \begin{eqnarray}
 \label{s-u symmetric ansatz}
   F(s,t) &\sim & -C' s\exp {(-i\pi /2)} (\ln (s/s_0) -i\pi/2)^\gamma \nonumber\\
   &\times& \phi (t(\ln (s/s_0) -i\pi/2)^\beta)
 \end{eqnarray}
Expanding in powers of $1/ \ln {s}$ at fixed $\tau$ we get, 
\begin{eqnarray}
&\frac{Im F(s,t)}{Im F(s,0)}& \rightarrow \phi (\tau);\\
\label{scaling2}
& \frac{Re F(s,t)}{Im F(s,0)}& \rightarrow \frac{\pi}{2 \ln {(s/s_0)}}\bigl( \gamma \phi (\tau) + \beta \tau \phi '(\tau)  \bigr),\\
&\frac{Re F(s,t)}{s}& \rightarrow (\pi/2) (\frac{\partial (Im F(s,t)/s) }{ \partial (\ln(s/s_0))});\\
\label{Re a(b,s)}
&Re a(b,s)& \rightarrow (\pi/2)\frac{\partial (Im a(b,s))}{\partial \ln(s/s_0)},
\end{eqnarray}
where, due to linearity, the last two equations also hold for a superposition of terms of the form (\ref{s-u symmetric ansatz}), e.g. $F^{(1)}+F^{(2)}$. Note that,
(i) $Re F(s,0)/Im F(s,0)$ agrees with the Khuri-Kinoshita theorem \cite{KK1965}, (ii)the case $\beta=\gamma=1$ agrees with Martin's geometrical scaling 
formula \cite{Martin1973} \cite{Martin1997}. 
(iii) When  $\sigma_{tot}\sim (\ln{s})^2 , \gamma=\beta=2$ ,
the AKM theorem and Auberson-Roy theorem \cite{AKM1971}\cite{Auberson-Roy} guarantee the scaling of $Im F(s,t)/Im F (s,0)$ with  $\phi (\tau)$ being  
an entire function of order half.The crucial new result is the formula (\ref{scaling2}) for $Re F(s,t)$ .In turn, 
this yields for the partial waves of $F^{(2)}$, if $b^2 Im a^{(2)}(b,s)\rightarrow 0$ for $b\rightarrow \infty$,
\begin{equation}
 Re\> a^{(2)}(b,s)\rightarrow\frac{-\pi}{2\ln (s/s_0)}b \frac{\partial}{\partial b}Im\> a^{(2)}(b,s) ,\>s\rightarrow \infty.
\end{equation}
However, in view of the slow approach to asymptotics , the formula (\ref{Re a(b,s)}) for $Re a(b,s)$ involving derivative over $\ln s$ is preferable for computations, 
as it holds also for $F^{(1)}+F^{(2)}$.

\begin{figure}[!] 
% \begin{center}
\includegraphics[width=1.0\columnwidth]{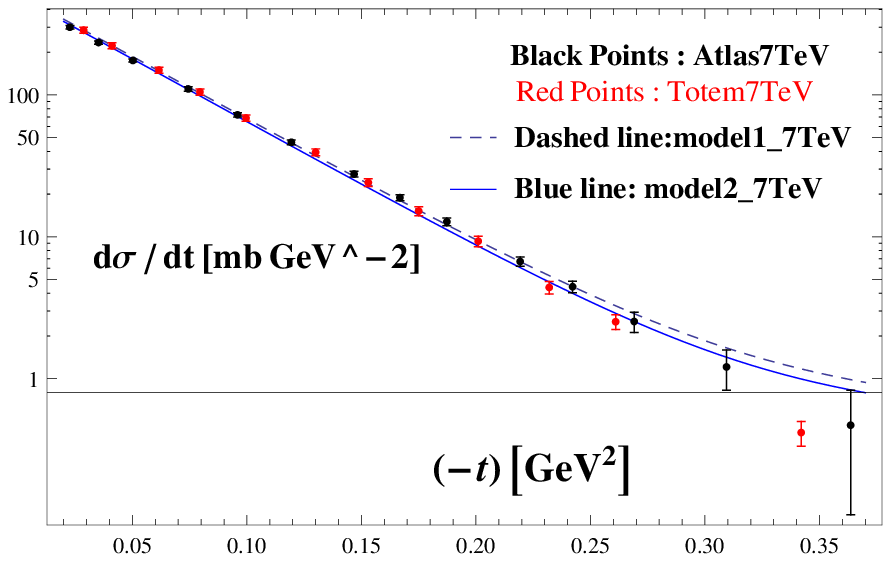}
\caption{Model predictions for pp elastic differential cross sections $d\sigma/
   dt  $   at  $7 TeV$  , 
with parameters $b_P = 3.8 GeV^{-2}, \alpha' = 
 0.07 GeV^{-2} $, 
forward slope  from Schegelsky - 
 Ryskin  fit \cite{Schegelsky-Ryskin} , input $\sigma_ {tot} $ from PDG (2005)\cite{PDG2005} (dashed curve), 
and input $\sigma_ {tot} $ from PDG (2013)\cite{PDG2013} (solid curve), 
show excellent agreement with experimental values from the Totem \cite{Totem1}\cite{Totem2}\cite{Totem3}\cite{Totem4}  and Atlas 
\cite{Atlas1}\cite{Atlas2}\cite{Atlas3}\cite{Atlas4} collaborations for $ | t | < 0.3 GeV^2 $.}
\label{Fig.11_Proton_quartic.eps}
% \end{center} 
\end{figure}

\begin{figure}[!] 
% \begin{center}
\includegraphics[width=1.0\columnwidth]{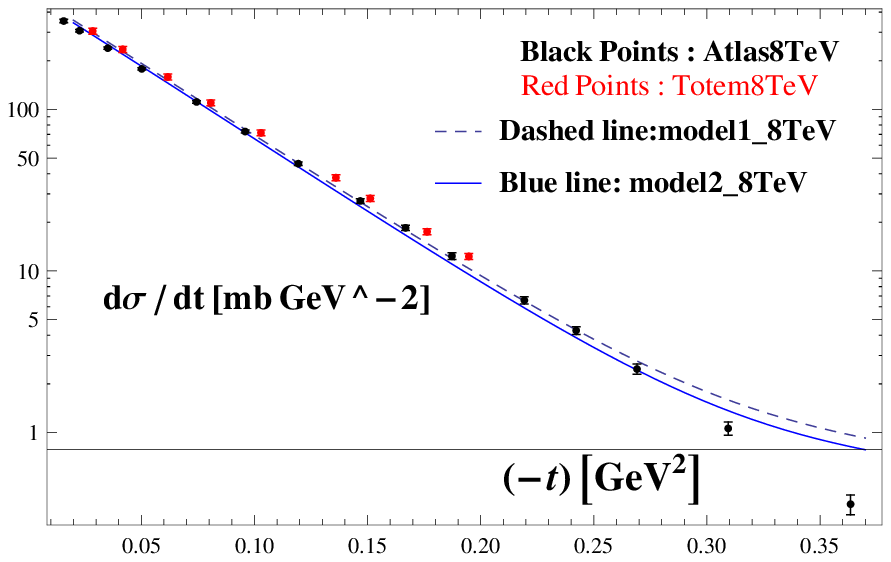}
\caption{Model predictions for pp elastic differential cross sections $d\sigma/
   dt  $   at  $8 TeV$  , 
with parameters $b_P = 3.8 GeV^{-2}, \alpha' = 
 0.07 GeV^{-2} $, 
forward slope  from Schegelsky - 
 Ryskin  fit \cite{Schegelsky-Ryskin} , input $\sigma_ {tot} $ from PDG (2005)\cite{PDG2005} (dashed curve), 
and input $\sigma_ {tot}$ from PDG (2013)\cite{PDG2013} (solid curve), 
show excellent agreement with experimental values from the Totem \cite{Totem1}\cite{Totem2}\cite{Totem3}\cite{Totem4}  and Atlas 
\cite{Atlas1}\cite{Atlas2}\cite{Atlas3}\cite{Atlas4} collaborations for $ | t | < 0.3 GeV^2 $.}
\label{Fig.12_Proton_quartic.eps}
% \end{center} 
\end{figure}

 \begin {figure}[!]
\includegraphics[width=1.0\columnwidth]{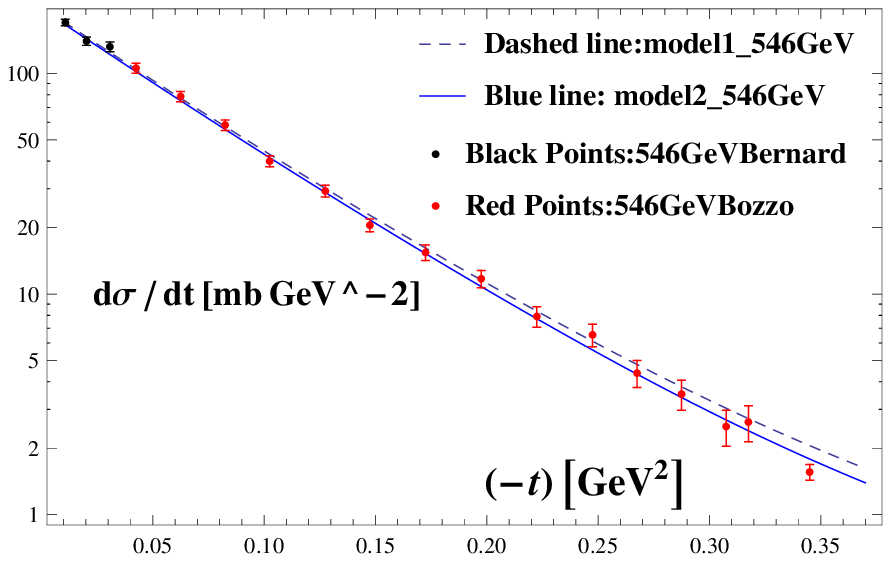}
\caption {Model predictions for $\bar{p}p$ elastic differential cross 
sections $d\sigma/dt $  at $546 GeV$,
   with parameters $b_P = 3.8 GeV^{-2}, \alpha' = 0.07 GeV^{-2} $, 
  forward slope from Schegelsky - 
    Ryskin  fit \cite{Schegelsky-Ryskin} , 
  input $\sigma_ {tot} $ from PDG (2005)\cite{PDG2005} (dashed curve), 
  and input $\sigma_ {tot}  $ from PDG (2013)\cite{PDG2013} (solid curve), 
  show good agreement with experimental values from UA4 collaborations, D. Bernad et al\cite{Bernard} and 
  M. Bozzo et al \cite{Bozzo} for $ | t | < 0.3 GeV^2 $.}
\label {Fig.15_Proton_quartic.eps}
\end {figure}

 \begin {figure}[!]
\includegraphics[width=1.0\columnwidth]{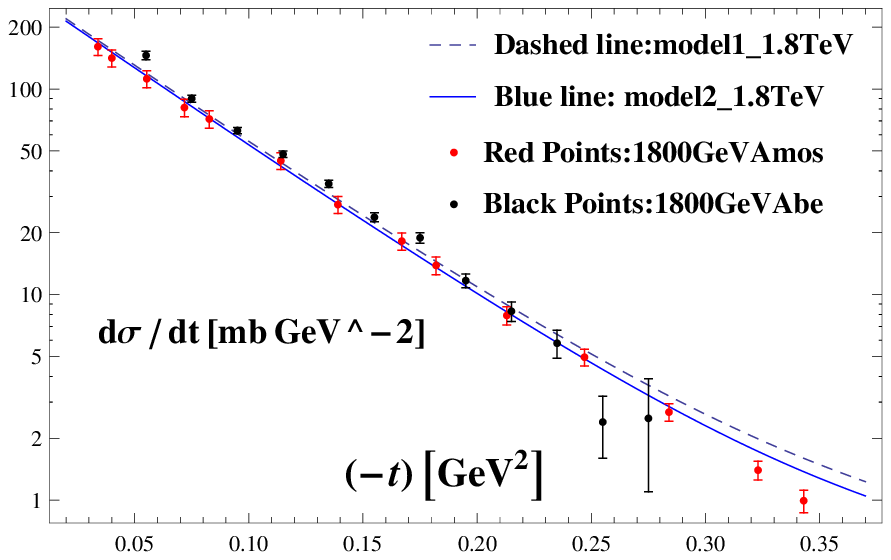}
\caption {Model predictions for $\bar{p}p$ elastic differential cross 
sections $d\sigma/dt $ at $1800 GeV$,
   with parameters $b_P = 3.8 GeV^{-2}, \alpha' = 0.07 GeV^{-2} $, 
  forward slope from Schegelsky - 
    Ryskin  fit \cite{Schegelsky-Ryskin} , 
  input $\sigma_ {tot}$ from PDG (2005)\cite{PDG2005} (dashed curve), 
  and input $\sigma_ {tot} $ from PDG (2013)\cite{PDG2013} (solid curve), 
  show good agreement with experimental values from Amos et al \cite{Amos} and 
 Abe et al \cite{Abe} for $ | t | < 0.3 GeV^2 $.}
\label {Fig.16_Proton_quartic.eps}
\end {figure}

\begin {figure}[h]
\includegraphics[width =1.0\columnwidth]{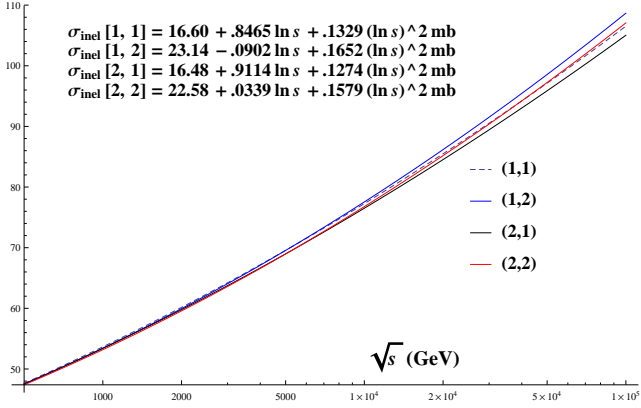} 
 \caption { Plots of $pp$ inelastic cross sections $\sigma_ {inel} (q, 
   M) $ computed from the model with $q = 
 1 $ and $q = 2 $ signifying inputs 
of $\sigma_{total} (PDG-2005)$ \cite{PDG2005} and  $\sigma_{total} (PDG-2013) $\cite{PDG2013} respectively and $M = 
 1 $ and $M = 
  2 $ signifying inputs of Okorokov \cite{Okorokov} and Schegelsky - 
   Ryskin \cite{Schegelsky-Ryskin} slopes respectively.Input Pomeron parameters are $b_ P = 
   3.8 GeV^{-2}, \alpha' = 0.07 GeV^{-2} $.Three parameter fits to these inelastic cross sections are also shown.
  }
\label {Fig.14_Proton_quartic.eps} 
\end {figure}

{\bf The total amplitude.} Consistent with (\ref{scaling2}) for $\gamma=\beta=2$, i.e. $ \tau= t (\ln{|s/s_0|})^2$,I adopt the ansatz,
\begin{equation}
\frac{Re F^{(2)}(s,t)}{ImF^{(2)}(s,0)}= \frac{\pi}{\ln (s/s_0)} \frac{d}{d\tau} \big(\tau \frac{Im F^{(2)}(s,t)}{ImF^{(2)}(s,0)} \big).
\end{equation}
For simplicity, I choose the scale factor $s_0$ to be the same as in the PDG(2005)\cite{PDG2005} fit for $pp$ total cross section,$ \sqrt{s_0}=5.38 GeV.$
Substituting the expression for $ Im F^{(2)}(s,t)$ I obtain, 
\begin{eqnarray}
\label{grey disk}
 \frac{16\pi}{k\sqrt{s}} F^{(2)}(s,t)=\sigma ^{(2)}_{tot} (s) \big[\frac{\pi}{\ln (s/s_0)}\nonumber\\
 \times \frac{8J_2(qR(s))-16J_4( qR(s))}{q^2R^2(s)} +i \frac{48J_3(qR(s))}{(qR(s))^3} \big].
\end{eqnarray}
The total amplitude $F(s,t)=F^{(1)}(s,t)+F^{(2)}(s,t)$ is now completely specified (analytically) by adding $F^{(1)}(s,t)$ 
given by (\ref{Pomeron}).The important parameter $R^2(s)$ is 
determined from the experimental slope parameter $B(s)=(d/dt)\big(\ln {d \sigma/dt} \big)|_{t=0},$
if the Pomeron parameters $b_P,\alpha'$ are known,
\begin{eqnarray}
& & R^2(s)\big(\epsilon (s)\sigma ^{(2)}_{tot} (s)^2 +\frac{1}{2}\sigma ^{(2)}_{tot} (s) \sigma _{tot} (s) \big) \nonumber \\
 &=&4B(s)\big(\epsilon (s)\sigma ^{(2)}_{tot} (s)^2 +\sigma _{tot} (s) ^2 \big)\nonumber \\
 &-&\sigma ^{(1)}_{tot}  \sigma _{tot} (s)D^2(s)-4 \pi \alpha '\sqrt{\epsilon (s)}\sigma ^{(1)}_{tot} \sigma ^{(2)}_{tot} (s),
\end{eqnarray}
where, we denote $\sqrt{\epsilon(s)} \equiv \pi/\ln {(s/s_0)}.$
For the experimental slope parameter I shall use the fits $B(M,s)$ to all $pp$ data , with $M=1,2$,  $ B(1,s)$ by Okorokov \cite{Okorokov} and $B(2,s)$ by 
Schegelsky-Ryskin \cite{Schegelsky-Ryskin} ,
\begin{eqnarray}
 B (1, s) &=& 8.81 + 0.396 ln s + 0.013 (ln s)^2 \>GeV^{-2},\nonumber \\
 B (2, s) &=& 11.03 + 0.0286 (ln s)^2 \>GeV^{-2},
\end{eqnarray}
where $\sqrt {s}$ is in $GeV$ units. For $pp,\bar{p}p$ total cross sections I use the PDG fits of (2005) and (2013),
\begin{eqnarray}
 \sigma^{(2005)}_{tot}(s)&=&35.63+0.308 \big( ln (\frac{s}{28.94})\big)^2 \>mb \nonumber\\
 \sigma^{(2013)}_{tot}(s)&=&33.73+0.2838 \big( ln (\frac{s}{15.618})\big)^2 \>mb .
\end{eqnarray}

{\bf Elastic and inelastic cross sections.} 
The integrals over impact parameter needed to calculate $\sigma_{el}$ can be done exactly. We obtain,
\begin{eqnarray}
&& \sigma_{el} (s)=(\pi/2)C^2(s) D^2(s)(2+(\beta'(s))^2)\nonumber\\
&&+4\pi R^2(s)E^2(s)(3+2\epsilon (s))/15 \nonumber\\
&&+2\pi R^2(s)C(s)E(s)\delta ^{-3}(s) \big[\> \exp {(-2\delta(s))}\nonumber\\
&&\times (-1+2\beta '(s)\sqrt{\epsilon(s)}(2\delta ^2(s) +3\delta (s)+2)\>) +\nonumber \\
&&(2\beta' (s)\sqrt{\epsilon(s)}(\delta (s)-2) + 2\delta ^2(s) -2\delta (s)+1)   \big],\nonumber\\
&& \delta(s)\equiv R^2(s)/D^2(s),\>\beta'(s) \equiv 4\pi \alpha '/D^2(s).
\end{eqnarray}

\begin{table}[!]
% \begin{center}
 \includegraphics[width=1.0\columnwidth]{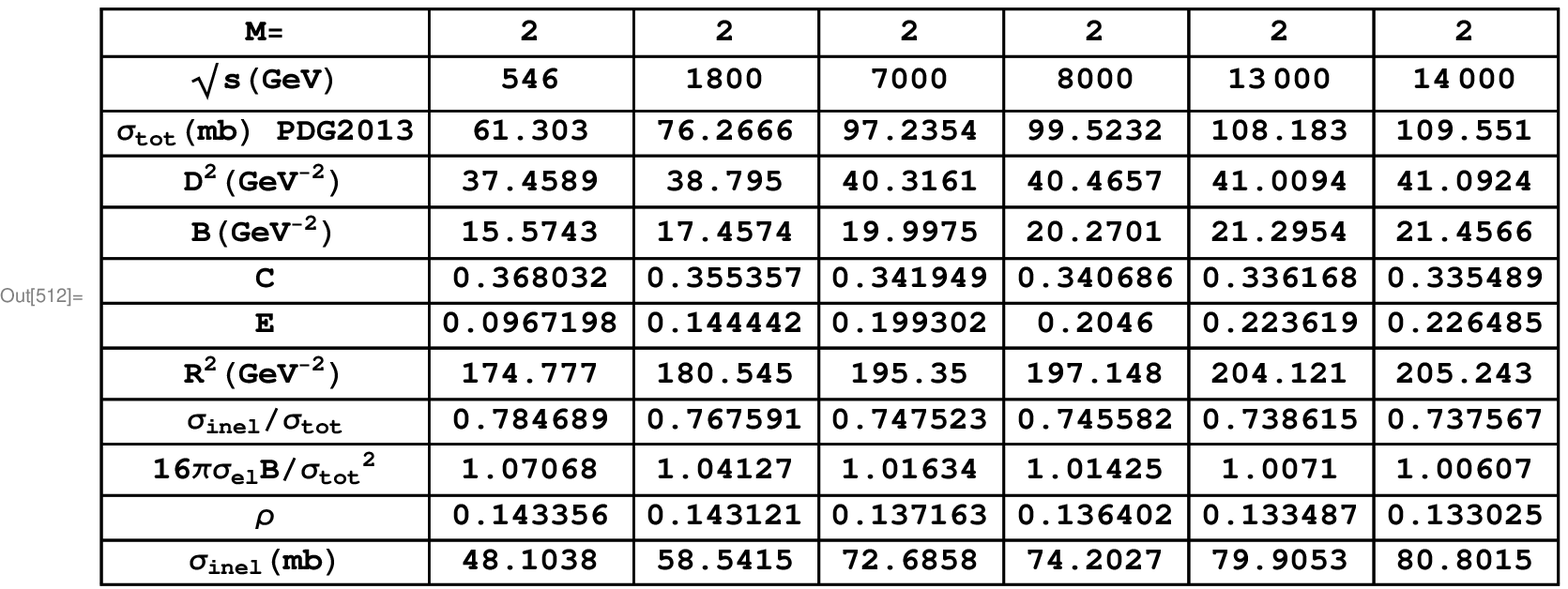}
\caption{Detailed results at 546 GeV,1.8 TeV, 7 TeV, 
8 TeV ,13 TeV and 14 TeV from the model using inputs $b_P = 3.8, \alpha' = 
.07 GeV^{-2}$, PDG 2013 values of $\sigma_ {tot} (pp) $ \cite{PDG2013}, and Schegelsky-Ryskin  
extrapolations $ (M = 2, i.e.B = B(2,s)) $\cite{Schegelsky-Ryskin} for forward slopes. 
The output parameters $C$ and $E$ show explicitly that inelastic 
unitarity is obeyed.The output values of $R^2 $ show a 
slowly expanding size of the proton with increasing energy.The 
output results for $ \sigma_ {inel} /\sigma_ {tot}$  , 
$16\pi \sigma_ {el} B/\sigma_ {tot}^2 $, and $\rho=ReF(s,t=0)/Im F(s,t=0)$, which would be 
1/2, 1 and 0 respectively in the black disk limit, give quantitative 
measures for  deviations from that limit.The output $\rho$ agrees with available experiments \cite{Compete1}\cite{Compete2}. 
The output values of $\sigma_ {inel} $ agree within errors with Totem results \cite{Totem1}\cite{Totem2}
   \cite{Totem3}\cite{Totem4} and Atlas results \cite{Atlas1}\cite{Atlas2}\cite{Atlas3}\cite{Atlas4} for $pp$ scattering at 7 TeV and 8 TeV, 
  and with the results of \cite{Abe} for $\bar{p}p$ scattering at 546 GeV and 1800 GeV.   
   Model  predictions at higher energies can be tested in future 
experiments.}
 \label{Table16_Proton_quartic.eps} 
% \end{center} 
\end{table}

\begin{table}[!]
% \begin{center}
 \includegraphics[width=1.0\columnwidth]{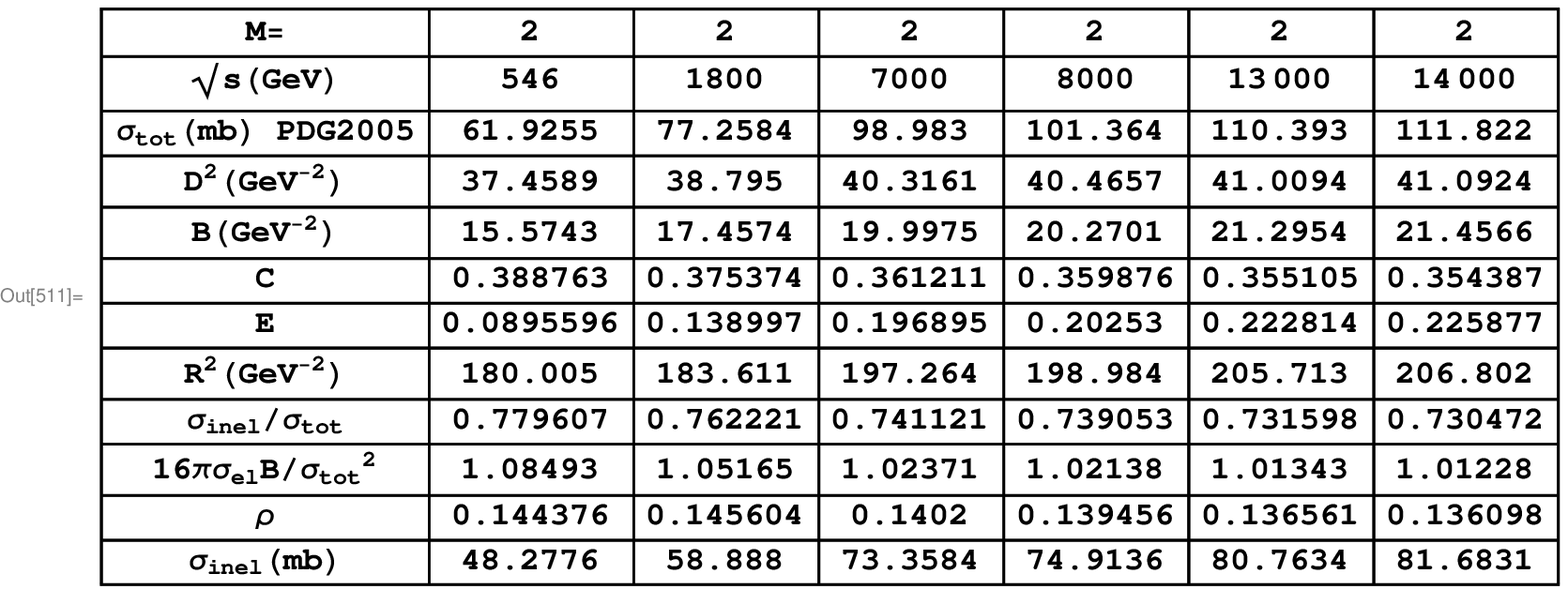}
\caption{Same as Table (\ref{Table16_Proton_quartic.eps} ) , but for input $\sigma_ {tot} (PDG - 
    2005)$. Comparison shows that the predicted inelastic cross section 
at 7 TeV  (8 TeV) increases by about 0.7 mb, 
   when the input $\sigma_ {tot}$ increases  by 1.8 mb ( 
  1.9 mb). }
 \label{Table17_Proton_quartic.eps} 
% \end{center} 
\end{table}

{\noindent \bf Predictions of the model versus experimental data for $pp$ and $\bar{p}p$ scattering.}

{\bf Differential cross sections}.Remarkably, a single pair of values of the Pomeron parameters $b_P, \alpha '$ ,
\begin{equation}
\label{Pomeron parameters}
 b_ P = 3.8 GeV^{-2}, \alpha'= 0.07 GeV^{-2}. 
\end{equation}
gives very good agreement of model predictions in the entire range $ | t | < 0.3 GeV^2 $ with the experimental Totem \cite{Totem1}\cite{Totem2}\cite{Totem3}\cite{Totem4} 
and Atlas \cite{Atlas1}\cite{Atlas2}\cite{Atlas3}\cite{Atlas4} $pp$ differential cross sections  at $7 TeV$ and $8 TeV$ 
, experimental  $\bar{p}p$ differential cross sections at $546 GeV$ from UA4 collaborations, D. Bernard et al\cite{Bernard} and 
  M. Bozzo et al \cite{Bozzo}, and at $1800 GeV$ from Amos et al \cite{Amos} and Abe et al \cite{Abe}.(See also the compilation in \cite{Cudell2006}]. This agreement 
  is independent of the choice between PDG(2005) and PDG(2013)  total cross sections, and the choice between slopes $B(1,s)$ and  $B(2,s)$. 
  We exhibit this in Figs.(\ref{Fig.11_Proton_quartic.eps} ,\ref{Fig.12_Proton_quartic.eps},\ref{Fig.15_Proton_quartic.eps},
  \ref{Fig.16_Proton_quartic.eps}) for forward slope choice $B = B(2,s)$  \cite{Schegelsky-Ryskin} and the two choices of total cross sections 
PDG (2005)\cite{PDG2005} (dashed curve), and PDG (2013)\cite{PDG2013} (solid curve).(Differential cross sections for $(-t)>0.3 GeV^2$ are not used in 
determination of Pomeron parameters $b_P,\alpha '$ as they make negligible 
contributions to $\sigma_{el}$  in this energy range; e.g. in this model, about $0.2 mb$ at $7 TeV$ and $8 TeV$. )

 For the choice $B = B(2,s)$  \cite{Schegelsky-Ryskin} and PDG (2013)\cite{PDG2013} total cross sections, we give below three parameter fits to predicted 
differential cross sections in this range of $t$ at c.m. energies upto $14\> TeV$,
\begin{eqnarray}
 &\ln ((d\sigma /dt)/(d\sigma /dt) _ {t = 0})  \nonumber \\
 &  = 19.5 t - 11.9 t^2 + 43.5 (-t)^3,\> 7 TeV \nonumber\\
 & = 19.7 t - 13.2 t^2 + 47.3 (-t)^3 ,\> 8 TeV\nonumber\\
 & =20.5 t - 19.2 t^2 + 64.2 (-t)^3, \>13 TeV \nonumber\\
 & =  20.6 t - 20.3 t^2 + 67.2 (-t)^3 .\> 14 TeV
\end{eqnarray}
  
for ready comparisons with existing and future data.

{\bf Inelastic cross sections} . Fig.(\ref{Fig.14_Proton_quartic.eps} ) 
depicts the  predicted inelastic cross sections up to $100 TeV$ and their asymptotic fits. Tables (\ref{Table16_Proton_quartic.eps} ) and 
(\ref{Table17_Proton_quartic.eps} ) give model 
parameters and detailed predictions from $546 GeV$ to $14 TeV $, with input total cross sections $PDG2013$ and $PDG2005$ respectively.
The predicted $\rho=Re F(s,t)/Im F(s,t)|_{t=0} $ and the predicted inelastic cross sections (e.g. for input total cross section $PDG2013$ ,
$\rho=0.136,\>\sigma_{inel}=74.2\>mb,$ at $8TeV$ )
are very close to available experimental values \cite{Compete1}\cite{Compete2},\cite{Totem1}\cite{Totem2}
   \cite{Totem3}\cite{Totem4}\cite{Atlas1}\cite{Atlas2}\cite{Atlas3}\cite{Atlas4}
   . The predicted inelastic cross 
sections are fairly robust, changing by less than 0.5 mb in the range $(7TeV,14 TeV)$ 
when the slope parameter is changed from $B(1,s)$ to $B(2,s)$ and by less than $1 mb$
when the input $\sigma_{tot}$ is changed from PDG (2005) to PDG (2013) .Model results give $ \partial \sigma_{inel}/\partial B \sim 1.07 mb\>GeV^2,
\>\partial \sigma_{inel}/\partial \sigma_{tot} \sim 0.46$, and using input errors of PDG2013 fits, and $\delta B \sim 0.3 GeV^{-2}$ 
upto $100 TeV$ \cite{Schegelsky-Ryskin}, 
I have the error estimate ,$ \delta \sigma_{inel} \sim .47+.0021\big( ln {(s/15.618)}\big)^2 \>mb. $

  In the c.m. energy range from $0.5 TeV$ to $100 TeV $,the model parameters are very well approximated by the following fits.
\begin{eqnarray}
\label{model parameters q=1}
&&Input \> \sigma^{(2005)}_{tot}(s):\nonumber\\
&&M=1: E(s)= 0.987849 - 20.3797 /x + 113.797 /x^2\nonumber\\
&&M=1: R^2(s)=241.078 - 9.20435 x + 0.375387 x^2 \nonumber\\
&&M=2:E(s)= 0.861023 - 16.7296 /x + 88.3041 /x^2\nonumber\\
&&M=2: R^2(s)=245.408 - 11.3716 x + 0.487702 x^2 
\end{eqnarray}

\begin{eqnarray}
\label{model parameters q=2}
&&Input \> \sigma^{(2013)}_{tot}(s):\nonumber\\
&&M=1:E(s)=0.936736 - 18.91 /x + 104.505 /x^2 \nonumber\\
&&M=1: R^2(s)= 214.735 - 6.85598 x + 0.320973 x^2 \nonumber\\
&&M=2:E(s)=0.812299 - 15.3352 /x + 79.6064 /x^2 \nonumber\\
&&M=2: R^2(s)= 220.921 - 9.20272 x + 0.437436 x^2 
\end{eqnarray}
where,$ x\equiv \ln{s}.$

Remarkably, fits for input $\sigma^{(2005)}_{tot}(s)$ show that the choice $M=1$  gives
$E(s)$ which is barely below the unitarity limit for $s\rightarrow \infty.$ The inelastic cross section fits in Figure \ref{Fig.14_Proton_quartic.eps} yield ,
 \begin{eqnarray}
&Input\>\sigma^{(2013)}_{tot}(s):\nonumber\\
&M=1:\frac{\sigma_{inel}}{\sigma_{tot}} \rightarrow 0.449 ;M=2:\frac{\sigma_{inel}}{\sigma_{tot}} \rightarrow 0.556\nonumber\\
&Input\>\sigma^{(2005)}_{tot}(s): \nonumber\\
&M=1:\frac{\sigma_{inel}}{\sigma_{tot}} \rightarrow 0.431 ;M=2:\frac{\sigma_{inel}}{\sigma_{tot}} \rightarrow 0.536 
\end{eqnarray}

These results are close to the black disk value of $1/2$ favoured by BH \cite{BH1}\cite{BH2}.Recent detailed 
analysis of high energy data \cite{Fagundes} concluded that, although consistent with experimental data, the black disk 
does not represent an unique solution.

 {\bf Phenomenological lowest t-channel singularity.} If continued to complex $t$, $|F(s,t)|$ given by this model is bounded 
 by $Const. s^2$ for $s\rightarrow \infty $ and 
\begin{equation}
\label{glue ball mass}
 |t|< t_1=min [(1/\alpha'),lim_{s \rightarrow \infty} (\ln{s}/R(s))^2 ].
\end{equation}
Jin and Martin \cite{Jin-Martin1964} proved that for $|t|<t_0 $, where $t_0$ is the lowest 
 $t$-channel singularity,twice subtracted dispersion relations in $s$ hold.
Hence $t_1$ may be thought of as a phenomenological lowest $t-channel$ singularity.  Using the formulae for 
$R^2(s)$ given above,
\begin{eqnarray}
&& Input \>\sigma^{(2013)}_{tot}(s):\nonumber\\
&M&=1:\sqrt{t_1}=1.765\> GeV;M=2:\sqrt{t_1}=1.512\> GeV;\nonumber\\
&&Input \> \sigma^{(2005)}_{tot}(s): \nonumber\\
&M&=1:\sqrt{t_1}=1.632\> GeV;M=2:\sqrt{t_1}=1.432\> GeV.\nonumber
\end{eqnarray}

Our $\sqrt{t_1}\sim 1.4\> - 1.8 GeV $ is reminiscent of , but different from the 
glue-ball mass of BH \cite{BH1}\cite{BH2}. Given the instability of analytic continuations, its main function is to suggest that the 
usual Lukaszuk-Martin bound  \cite{Lukaszuk} is quantitatively poor as it assumes lack of $t-$channel singularities only upto $ 4 m_\pi ^2$ 
which is much smaller than $t_1$.

{\bf Conclusion}. I presented an analytic formula for the high energy elastic amplitude $F(s,t)=F^{(1)}(s,t)+F^{(2)}(s,t)$ given by Eqns. 
(\ref{Pomeron},\ref{grey disk}) for $\sqrt{s}>100 GeV $, exhibiting Froissart bound saturation, AKM scaling 
\cite{AKM1971}\cite{Auberson-Roy}, inelastic unitarity , predicting differential cross sections for $(-t)<0.3 GeV^2$ and 
total inelastic cross sections, at $546 GeV$, $1800 GeV$, 
$7 TeV$ and $8 TeV$ in agreement with experimental results, as well as the  
real parts and inelastic cross sections upto $100 TeV$. An 'effective' t-channel singularity 
at  $\sqrt{t} \sim 1.4-1.8 GeV$ is suggested 
by analytic continuation to positive $t$.
Detailed tables and graphs of model parameters, real parts and cross sections upto $100\>TeV$ will be published separately. The 
'grey disk' component could be generalized using a smoother impact parameter cut-off, i.e. $n>1$ in Eqn. (\ref{n}).

{\bf Acknowledgements}.  I presented an earlier version with a black disk second component   
in 2015 to Andr\'e Martin and T.T. Wu at CERN; their insistence that a sharp impact parameter cut-off is too 'brutal'  
led to the black disk being replaced by the grey disk. I thank G. Auberson for remarks concerning instability of analytic continuation, D. Atkinson, G. Mahoux 
and V. Singh for helpful comments on the manuscript; I also thank Gilberto Colangelo and Heiri Leutwyler for  very helpful discussions, and a 
seminar invitation  at Univ. of Bern, and Irinel Caprini and Juerg Gasser for discussions  
on a very stimulating  ansatz for high energy pion-pion scattering \cite{Caprini}. I thank the referees for the crucial suggestion of comparison with the 
latest differential cross section data and the Indian National Science Academy for an INSA senior 
scientist grant.

% May skip begin and end figure options for lectures
% Caption works in the (begin and end) figure environment only
%\begin{figure}[ht]
%\begin{center}
% Use LaTeX and DVItoPDF while using EPS files
%\includegraphics[width=.75\columnwidth]{position_density.eps}
% Use PDFLaTeX while using JPG, PNG, PDF image files
%\includegraphics[width=.75\columnwidth]{W3.jpg}
%\caption{Position density}
%\label{fig:pos_dens}
%\end{center}
%\end{figure}


\begin{thebibliography}{99}
\bibitem{AKM1971} G. Auberson,T. Kinoshita and A. Martin, Phys. Rev. D {\bf 3}, 3185 (1971).
\bibitem{Auberson-Roy} G. Auberson and S. M. Roy, Nucl. Phys.{\bf B117},322(1976).
\bibitem{PDG2005} S. Eidelman et al (Particle data Group) Phys. Lett.B{\bf 592},1(2004) and 
2005 partial update, http://pdg.lbl.gov/2005,Table 40.2.
\bibitem{PDG2013} J. Beringer et al (Particle data Group) Phys. Rev. D{\bf 86},010001(2012) and 
2013 partial update, http://pdg.lbl.gov/2013,Table 50.
\bibitem{Schegelsky-Ryskin} V. A.Schegelsky and M.G.Ryskin, Phys. Rev. D{\bf 85},094024 (2012).
\bibitem{Okorokov} V. A. Okorokov, arXiv:1501.01142 v2[hep-ph]23 May, 2015.
\bibitem{Totem1} Totem collaboration, G. Antchev et al, Europhys. Lett. {\bf 96},21002(2011) . 
\bibitem{Totem2} Totem collaboration, G. Antchev et al,Europhys. Lett.{\bf 101},21004(2013);CERN-PH-EP-2012-239. 
\bibitem{Totem3} Totem collaboration, G. Antchev et al,Phys. Rev. Lett. {\bf 111},012001 (2013).
\bibitem{Totem4} Totem collaboration, G. Antchev et al,;Nucl. Phys. B {\bf 899},527 (2015);arXiv:1503.08111[hep-ex].
\bibitem{CMS} CMS collaboration, Phys. Lett. B{\bf 722},5 (2013); presentation at EPS-HEP conference,
Vienna,22-29 July (2015).
\bibitem{Atlas1} Atlas collaboration, Nature Comm.{\bf 2},463(2011).
\bibitem{Atlas2} Atlas collaboration,Nucl. Phys. B {\bf 889 },486 (2014);arXiv:1408.5778v2[hep-ex].
\bibitem{Atlas3} Atlas collaboration,ATLAS-CONF-2015-038.
\bibitem{Atlas4} Atlas collaboration,CERN-EP-2016-158(25 July 2016); arXiv:1607.06605v1[hep-ex] (submitted to Phys. Lett. B).
\bibitem{Alice} Alice collaboration, ArXiv:1208,4968(2012).
\bibitem{Pierre Auger} Pierre Auger collaboration, Phys. Rev. Lett. {\bf 109},062002(2012).
\bibitem{Froissart1961} M. Froissart, Phys. Rev. {\bf 123}, 1053 (1961).
\bibitem{Martin1966} A. Martin, Nuov. Cimen. {\bf 42}, 930 (1966).
\bibitem{Lukaszuk}L. Lukaszuk and A. Martin, Nuov. Cimen. {\bf 52A}, 122 (1967).
\bibitem{Roy1972} S. M. Roy, Phys. Reports, {\bf 5C}, 125 (1972).
\bibitem{Kupsch} J. Kupsch, Nuovo Cim. {\bf 71A},85 (1982).
\bibitem{Cheng-Wu1970} H. Cheng and T. T. Wu, Phys. Rev. Letters {\bf 24},1456 (1970).
\bibitem{BSW1} C. Bourrely, J. Soffer, and T. T. Wu, Phys. Rev. {\bf D19}, 3249 (1979).
\bibitem{BSW2} C. Bourrely, J. Soffer, and T. T. Wu,Nucl. Phys. 
{\bf B247}, 15 (1984).
\bibitem{BSW3} C. Bourrely, J. Soffer, and T. T. Wu,Z. Phys.C {\bf 37},369 (1988).
\bibitem{BSW4} C. Bourrely, J. Soffer, and T. T. Wu,Eur.Phys. J. C{\bf 28},97(2003).
\bibitem{BSW5} C. Bourrely, J. Soffer, and T. T. Wu,Eur.Phys. J. C{\bf 71},1061(2011).
\bibitem{Block1} M. M. Block et al, Phys. Rev. D{\bf 60},054024 (1999).
\bibitem{Block2} M. M. Block, Phys. Reports, {\bf 436}, 71 (2006).
\bibitem{Block3} M. M. Block and F. Halzen, Phys. Rev. D {\bf 83},077901(2011).
\bibitem{Islam} M. M. Islam et al,Int. J. Mod. Phys. A {\bf 21},1 (2006).
\bibitem{Martin2009} A. Martin, Phys. Rev. {\bf D80}, 065013 (2009).
\bibitem{WMRS} T. T. Wu, A. Martin, S. M. Roy and V. Singh, Phys. Rev. {\bf D84}, 025012 (2011).
\bibitem{Martin-Roy2014} A. Martin and S. M. Roy , Phys. Rev. D {\bf 89}, 045015 (2014).
\bibitem{Martin-Roy2015} A. Martin and S. M. Roy , Phys. Rev. D{\bf 91},076006 (2015).
\bibitem{AGN} R. F. Avila, P. Gauron and B. Nicolescu, Eur. Phys. J. C{\bf 49},581 (2007).
\bibitem{Martynov} E. Martynov and B. Nicolescu, Eur. Phys. J. C{\bf 56},57 (2008).
\bibitem{BH1} M. M. Block and F. Halzen, Phys. Rev. Lett. {\bf 107},212002 (2011).
\bibitem{BH2}Phys. Rev. D {\bf 86},051504(R) (2012).
\bibitem{KK1965} N. N. Khuri and T. Kinoshita, Phys. Rev.{\bf 137},B720 (1965).
\bibitem{Martin1973} A. Martin, Lett. Nuovo Cimento {\bf 7}, 811 (1973).
\bibitem{Martin1997} A. Martin,CERN-TH/97-23 (1997).
\bibitem{Bernard} D. Bernard et al, UA4 collaboration, Phys. Lett. B{\bf 198},583(1987).
\bibitem{Bozzo} M. Bozzo et al, UA4 collaboration, Phys. Lett. B{\bf 147},385(1984).
\bibitem{Amos} N. A. Amos et al,  Phys. Lett. B{\bf 247},127(1990).
\bibitem{Abe} F. Abe et al, Phys. Rev. D {\bf 50},5518(1994) and  Phys. Rev. D {\bf 50},5550(1994).
\bibitem{Cudell2006}J. R. Cudell, A. Lengyel, E. Martynov,Phys. Rev. D {\bf 73},034008 (2006) and 
arXiv: hep-ph/0511073.
\bibitem{Compete1} COMPETE Collaboration, J. R. Cudell et al, Phys. Rev. Lett. {\bf 89},201801 (2002).
\bibitem{Compete2} COMPETE Collaboration, J. R. Cudell et al,Phys. Rev. D {\bf 65} 074024 (2002).
\bibitem{Fagundes} D. A. Fagundes, M. J. Menon and P.V. R. G. Silva,, Nucl. Phys.{\bf A946},194 (2016).
\bibitem{Jin-Martin1964} Y. S. Jin and A. Martin, Phys. Rev. {\bf 135B}, 1375(1964).
%\bibitem{Greynal-de Rafael} D. Greynal and E. de Rafael, Phys. Rev. {\bf D88},034015 (2013).
%\bibitem{Greynal-de Rafael-Vulvert} D.Greynal, E. de Rafael and G.Vulvert, JHEP03,107(2014).
\bibitem{Caprini} I. Caprini,G. Colangelo, and H. Leutwyler, Eur.Phys.J. {\bf C72},1860,(2012)

\end{thebibliography}
\end{document}